\documentclass[journal=jacsat,manuscript=article]{achemso}

\usepackage{amsmath,amssymb}
\usepackage{siunitx}
\usepackage{float}
\usepackage{graphicx}
\usepackage{pgf}
\usepackage{tikz}
\usepackage[colorlinks=true, citecolor=blue, linkcolor=blue, urlcolor=blue, pdfborder={0 0 0}]{hyperref}
\usepackage[capitalise]{cleveref}

\usepackage{amsmath}
\usepackage{epstopdf}
\usepackage{dcolumn}
\usepackage{bm}
\usepackage{braket}
\usepackage{gensymb}
\usepackage{array}
\usepackage[utf8]{inputenc}
\usepackage[T1]{fontenc}
\usepackage{mathptmx}

\newcommand{\ketAgm}[1]{|#1^1A_g^-\rangle}
\newcommand{\ketAg}[1]{|#1^1A_g\rangle}
\newcommand{\Agm}[1]{#1^1A_g^-}
\newcommand{\Bup}[1]{#1^1B_u^+}

\newcommand{\Bum}[1]{#1^1B_u^-}
\newcommand{\ketBum}[1]{|#1^1B_u^-\rangle}
\newcommand{\ketBup}[1]{|#1^1B_u^+\rangle}

\newlength{\myfigwidth}

\title{Singlet Triplet-Pair Production and Possible Singlet-Fission in Carotenoids}

\author{Dilhan Manawadu}
\email{dilhan.manawadu@chem.ox.ac.uk}
\affiliation{Department of Chemistry, Physical and Theoretical
Chemistry Laboratory, University of Oxford, Oxford, OX1 3QZ, United
Kingdom\\Linacre College, University of Oxford, Oxford, OX1 3JA, United Kingdom}
\author{Darren J. Valentine}
\affiliation{Department of Chemistry, Physical and Theoretical
Chemistry Laboratory, University of Oxford, Oxford, OX1 3QZ, United
Kingdom\\Balliol College, University of Oxford, Oxford, OX1 3BJ, United
Kingdom}
\author{Max Marcus}
\affiliation{Department of Chemistry, Physical and Theoretical
Chemistry Laboratory, University of Oxford, Oxford, OX1 3QZ, United
Kingdom}
\author{William Barford}
\email{william.barford@chem.ox.ac.uk}
\affiliation{Department of Chemistry, Physical and Theoretical
Chemistry Laboratory, University of Oxford, Oxford, OX1 3QZ, United
Kingdom}

\begin{document}

\begin{abstract}
Internal conversion from the photoexcited state to a correlated singlet triplet-pair state is believed to be the precursor of singlet fission in carotenoids. We present numerical simulations of this process using a $\pi$-electron model that fully accounts for electron-electron interactions and electron-nuclear coupling. The time-evolution of the electrons is determined rigorously using the time-dependent density matrix renormalization method, while the nuclei are evolved via the Ehrenfest equations of motion. We apply this to zeaxanthin, a carotenoid chain with 22 conjugated carbon atoms (i.e., 11 double bonds). We show that the internal conversion of the photoexcited state to the singlet triplet-pair state occurs adiabatically via an avoided crossing within 100 fs and we predict a yield of $\sim 50\%$
We further discuss whether this singlet triplet-pair state will undergo exothermic versus endothermic intra or inter chain singlet fission.

\end{abstract}

\vfill\pagebreak


The exotic electronic states of polyenes have been of abiding interest for nearly 50 years.\cite{Hudson72,Schulten72,Hayden86,Tavan87,Book} Their fascinating properties arise because electron-electron (e-e) interactions and electron-nuclear (e-n) coupling are significantly enhanced in quasi-one dimensional systems. One of the consequences of these interactions is that the lowest-energy excited singlet state is the non-emissive $\Agm{2}$ state (labelled $S_1$) that has significant correlated triplet-pair (or bimagnon) character. In contrast, the optically excited $\Bup{1}$ state (labelled $S_2$) has  correlated electron-hole (or excitonic) character, and which -- in the absence of e-e interactions and e-n coupling -- would lie energetically below the $\Agm{2}$ state.
The energetic reversal of the bright ($S_2$) and dark ($S_1$) states has various photophysical consequences. For example, it explains the non-emissive properties of linear polyenes, it is responsible for the photoprotection properties of carotentoids in light harvesting complexes, and it is thought to be the cause of singlet fission in polyene-type systems\cite{Kraabel98,Lanzani99,Musser13,Kasai15,Busby15,Huynh17,Musser19,Sanders19}.

Singlet fission  is a process by which a photoexcited state dissociates into two non-geminate triplets. In carotenoids and polyenes, while uncertainty remains as to whether the final step is an intra or intermolecular processes, the first step is understood to be the internal conversion  of the photoexcited singlet into a correlated singlet triplet-pair  state.
In understanding the process of singlet fission, it is useful to recall how  a pair of triplets combine\cite{Musser19,Sanders19}, namely $T_1\otimes T_1 = S + T +Q$, where $T_1$ represents the lowest-energy triplet, and $S$, $T$ and $Q$  are the singlet, triplet and quintet `correlated triplet-pair' states, respectively.
Using the density matrix renormalization group (DMRG) method to solve the Pariser-Parr-Pople-Peierls (PPPP) model of $\pi$-conjugated systems, Valentine \emph{et al.}\cite{Valentine20} performed an extensive theoretical and computational study of  the triplet-pair states of polyene chains. They showed that the singlet triplet-pair  state forms a band of states, $\ketAgm{2},\ketBum{1},\ketAgm{3},\cdots$, each with different center-of-mass kinetic energies. In the long-chain limit, however, the kinetic energy of these low-energy states vanishes and their vertical energies converge to the same value. Importantly, this excited energy is $\sim 0.3$ eV below the vertical energy of the quintet triplet-pair  state.
Since it was also shown\cite{Valentine20} that this quintet is an unbound  pair of spin-correlated triplets, we can conclude that the triplet-pair binding energy in the  singlet triplet-pair is $\sim 0.3$ eV.
\footnote{A similar conclusion concerning the binding energies of correlated triplet-pairs was made by Taffet \emph{et al.}\cite{Taffet20}.}
In addition, the vertical and relaxed energies of these low-energy singlet triplet-pair  states lie below the vertical and relaxed energies of $S_2$.

This picture becomes more complicated and interesting when we consider carotenoid chain sizes (i.e., $N = 14 - 26$), as now the center-of-mass kinetic energy plays a role in the relative energetic ordering. In particular, it was shown in Ref.\ \cite{Valentine20} that for all chain lengths the vertical and relaxed $\Agm{2}$ energies  lie below the corresponding $\Bup{1}$ energies. In contrast, while the $\Bum{1}$ relaxed energy  is lower than the $\Bup{1}$ relaxed energy, its vertical energy is higher than the $\Bup{1}$ vertical energy for $N < 22$ C-atoms.  Similarly, the relaxed $\Agm{3}$ energy  lies lower than the relaxed  $\Bup{1}$ energy for $N > 26$, while its vertical energy is higher for $N < 42$. These energetic orderings therefore imply that for certain chain lengths a vertical excitation to the $\Bup{1}$ state will be followed by fast internal conversion to either the $\Bum{1}$ or $\Agm{3}$ states.
In addition, the relaxed $\Bum{1}$ energy  is lower than the relaxed quintet energy and \textit{vice versa} for the $\Agm{3}$  state, while the relaxed energies of both the  $\Bum{1}$ and $\Agm{3}$  states is more than twice the energy of the relaxed triplet.
Thus, internal conversion  to the $\Bum{1}$ or $\Agm{3}$  states implies potentially endothermic or exothermic intramolecular singlet fission, respectively, and  exothermic intermolecular singlet fission in both cases.

These theoretical results (obtained using the Chandross-Mazumdar\cite{Chandross97} parametrization of the PPP model) are qualitatively consistent with the experimental observations on carotenoids summarized in Fig.\ 1 of Ref\cite{Rondonuwu03}, with the difference being that experimentally the cross-over in $\Agm{3}$ and $\Bup{1}$ relaxed energies occurs at 20 sites (10 double bonds) rather than 26 sites.

In this work we investigate the internal conversion  from the photoexcited singlet, $S_2$, to the correlated singlet triplet-pair states in carotenoids. We perform rigorous dynamical simulations using a realistic model of $\pi$-electron conjugated systems that incorporates the key features of electron-electron repulsion and electron-nuclear coupling. The quantum system describing the electronic degrees of freedom is evolved via the time-dependent Schr\"odinger equation using the time-dependent DMRG (t-DMRG) method. t-DMRG is a very accurate method for simulating dynamics in highly correlated one-dimensional quantum systems.\cite{White04} The nuclear degrees of freedom are treated classically via the Ehrenfest equations of motion. The computational methods are described in the SI.

As t-DMRG is conveniently  implemented with only on-site and nearest neighbor Coulomb interactions, in this investigation the $\pi$-electron system is described by the extended Hubbard (or UV) model, defined by
\begin{equation}
\hat{H}_{\textrm{UV}} = -2\beta \sum_{n=1}^{N-1}  \hat{T}_n + U \sum_{n=1}^N  \big( \hat{N}_{n \uparrow} - \frac{1}{2} \big) \big( \hat{N}_{n \downarrow} - \frac{1}{2} \big)
+ \frac{1}{2} \sum_{n=1}^{N-1} V \big( \hat{N}_n -1 \big) \big( \hat{N}_{n+1} -1 \big).
\end{equation}
Here, $\hat{T}_ n = \frac{1}{2}\sum_{\sigma} \left( c^{\dag}_{n , \sigma} c_{n + 1 , \sigma} + c^{\dag}_{n+ 1 , \sigma} c_{n  , \sigma} \right)$ is the bond order operator and $\hat{N}_n$ is the number operator. $N$ is the number of conjugated carbon-atoms ($N/2$ is the number of double bonds), $\beta$ is the hopping integral for a uniform chain, $U$ is the Coulomb interaction of two electrons in the same orbital and $V$ is the nearest-neighbor Coulomb repulsion. Since the UV model does not contain the long-range Coulomb terms of the PPP model, as described in the SI it is necessary to parametrize $U$ and $V$ to reproduce the predictions of Ref.\cite{Valentine20}.

The electrons couple to the nuclei via changes in the C-C bond length, specifically via
\begin{equation}
	\hat{H}_{\textrm{e-n}} =  2 \alpha \sum_{n=1}^{N-1}  \left( u_{n+1} - u_n \right) \hat{T}_n ,
\end{equation}
where $\alpha$ is the electron-nuclear coupling parameter and $u_n$ is the displacement of nucleus $n$ from its undistorted position.
Finally, the nuclear potential and kinetic energies are described
\begin{equation}
	\hat{H}_{\textrm{nuclear}} = \frac{K}{2} \sum_{n=1}^{N-1} \left( u_{n+1} - u_n \right)^2 + \frac{1}{2m}\sum_{n=1}^N p_n^2,
\end{equation}
where
$K$ is the nuclear spring constant and $m$ is the oscillator reduced mass.

\begin{figure}[h!]
    \centering
    \includegraphics[width=1.4\myfigwidth]{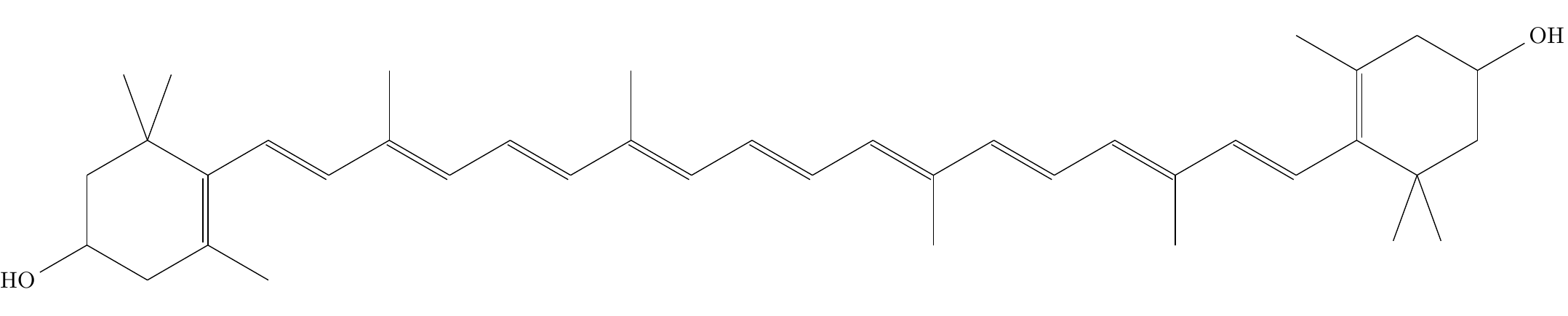}
    \caption{Structural formula of zeaxanthin.}
\label{Fig:4}
\end{figure}

The Hamiltonian $(\hat{H}_{\textrm{UV}}+\hat{H}_{\textrm{e-n}})$ is invariant under both a two-fold proper rotation (i.e., a $C_{2h}$ operation) and a particle-hole transformation (i.e., $(\hat{N}-1) \rightarrow -(\hat{N}-1)$), and so its eigenstates are labelled either  $A_g^{\pm}$ or $B_u^{\pm}$. Internal conversion from $S_2$ (i.e., the nominal $\Bup{1}$ state) to the triplet-pair singlets (with nominal negative particle-hole symmetry)\footnote{Here we adopt the particle-hole notation used in Ref.\ \cite{Valentine20}, which is typically used by the experimental community. It is the opposite definition to that used in Ref.\cite{Book,Barford01}.}  is achieved via an interaction that breaks particle-hole symmetry. Carotenoids naturally posses such an interaction because of their substituents, as described in the SI. This symmetry-breaking term is
\begin{equation}
	\hat{H}_{\textrm{SB}} =  \sum_{n=1}^{N} \epsilon_n (\hat{N}_n-1),
\end{equation}
which is odd under a particle-hole transformation.
Here, we investigate internal conversion in zeaxanthin, a carotenoid chain with $22$ conjugated C-atoms (and $11$ double bonds).
As shown in Fig.\ \ref{Fig:4}, zeaxanthin possess $C_{2h}$ symmetry and thus $\hat{H}_{\textrm{SB}}$ is even under this operation. From both energetic and symmetry considerations, therefore, only $\ketBup{1}$ to $\ketBum{1}$ internal conversion  is possible for this molecule.
We define the diabatic states as eigenstates of $(\hat{H}_{\textrm{UV}}+\hat{H}_{\textrm{e-n}})$, which  thus have   two-fold rotation and particle-hole symmetries. For our purposes the key  diabatic states are $\ketBum{1}$ and $\ketBup{1}$.
We define the adiabatic states as eigenstates of $(\hat{H}_{\textrm{UV}}+\hat{H}_{\textrm{e-n}} + \hat{H}_{\textrm{SB}})$, and thus these  states are linear combinations of  $\ketBup{1}$ and $\ketBum{1}$.

The initial state at time $t=0$ is defined via the operation of the dipole operator on the  ground state, i.e., $\ket{\Psi(t=0)} = \hat{\mu}\ket{\textrm{GS}}$. $\ket{\textrm{GS}}$ is the ground state obtained via static-DMRG\cite{White92} solutions of the  Hamiltonian $(\hat{H}_{\textrm{UV}}+\hat{H}_{\textrm{e-n}} + \hat{H}_{\textrm{SB}})$ coupled to a Hellmann-Feynman iterator to determine the equilibrium ground state geometry\cite{Barford01}.

\begin{figure}[h!]
\includegraphics[width=0.7\linewidth]{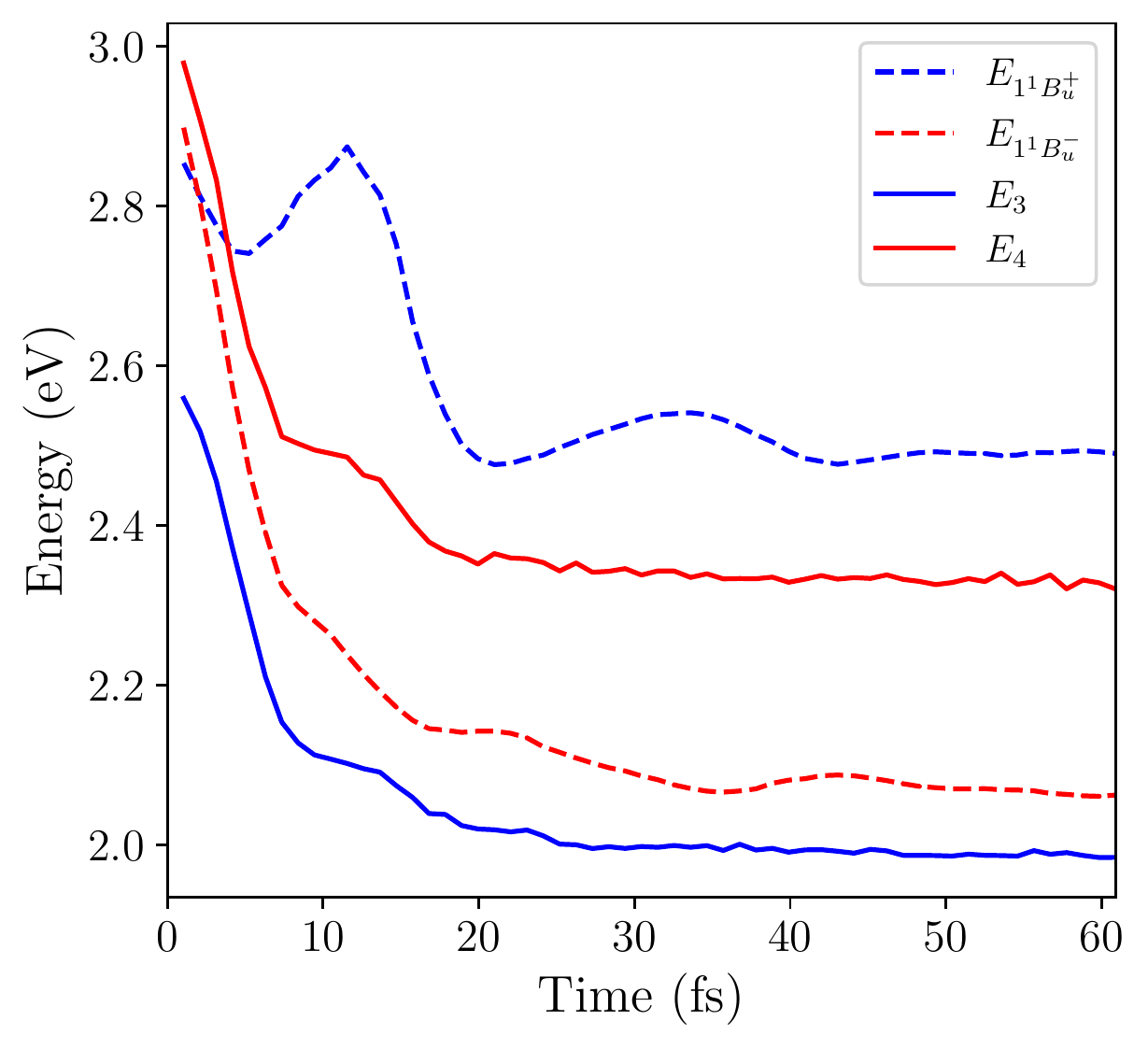}
\caption{Excitation energies of the diabatic $\Bum{1}$ 
and $\Bup{1}$ 
states, and the third
and fourth
adiabatic singlet states, ${\psi_3}$ and ${\psi_4}$, as a function of time.
These results are  for zeaxanthin, shown in Fig.\ 1.
The initial condition is $\ket {\Psi(0)} = \hat{\mu}\ket{\textrm{GS}}$. }
\label{Fig:1}
\end{figure}

We now describe our simulations for zeaxanthin. At the Franck-Condon point the forces exerted on the nuclei from the electrons in the excited state causes $\hat{H}_{\textrm{e-n}}$ to change, which is turn causes an evolution of the electronic and nuclear degrees of freedom.
As the system evolves there is a crossover  of the energies of the diabatic $\Bum{1}$ and $\Bup{1}$ states at  $\sim 2$ fs, as shown in Fig.\ \ref{Fig:1}.
The corresponding adiabatic energies (namely, the eigenvalues of the third and fourth singlet adiabatic states,  $\ket{\psi_3}$ and $\ket{\psi_4}$),\footnote{The first and second adiabatic states are $\ketAg{1}$ and $\ketAg{2}$, respectively.} however, exhibit an avoided crossing, because the coupling between the diabatic states, $\langle \Bup{1}|\hat{H}_{\textrm{SB}}|\Bum{1}\rangle$, remains non-zero throughout the evolution. (The avoided crossing is discussed in more detail in the SI).

\begin{figure}[h!]
\includegraphics[width=0.7\linewidth]{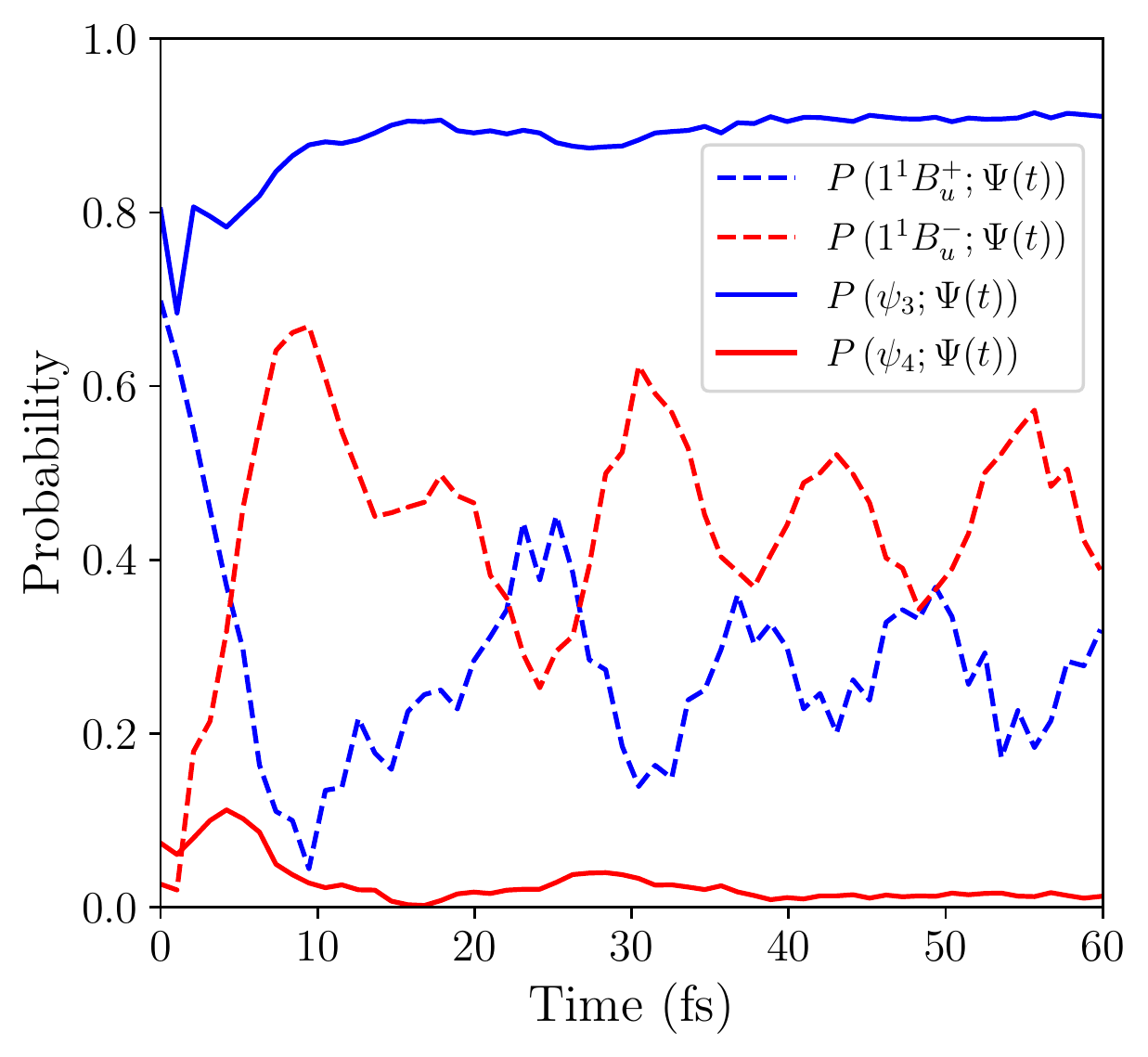}
\caption{Probabilities that $\ket{\Psi(t)}$ occupies the adiabatic states, $\ket{\psi_3}$ and $\ket{\psi_4}$, and the diabatic states, $\ketBup{1}$ and $\ketBum{1}$.
At time $t=0$, $\psi_3$ is the lowest-energy photoexcited state, $S_2$.
}
\label{Fig:2}
\end{figure}

The evolution of the  system described by $\ket{\Psi(t)}$ is illustrated in Fig.\ \ref{Fig:2}, which shows the probabilities that it occupies the third and fourth  adiabatic  singlet states.
The probability that $\ket{\Psi}$ initially occupies the lower adiabatic state $\ket{\psi_3}$ (i.e., that it is composed of  $S_2$) is $\sim 80\%$, because higher energy $^1B_u$ states (e.g., $\ket{\psi_4}$) are also excited by the dipole operator acting on the ground state. As the system evolves the probability that it occupies $\ket{\psi_3}$ increases  to $\sim 90\%$ within $\sim 10$ fs and then remains essentially constant, indicating that this is an adiabatic transition. The probability that the system occupies $\ket{\psi_4}$ rises to $\sim 10\%$ at the avoided crossing, before becoming negligible after $\sim 10$ fs.  As a consequence, the Ehrenfest approximation - which makes the erroneous assumption that  the nuclei  experience a mean force equal to the average from both adiabatic states\cite{Horsfield06, Tully12} - can be assumed to be largely valid here as only one state determines the  forces on the nuclei.

\begin{figure}[h!]
\includegraphics[width=0.7\linewidth]{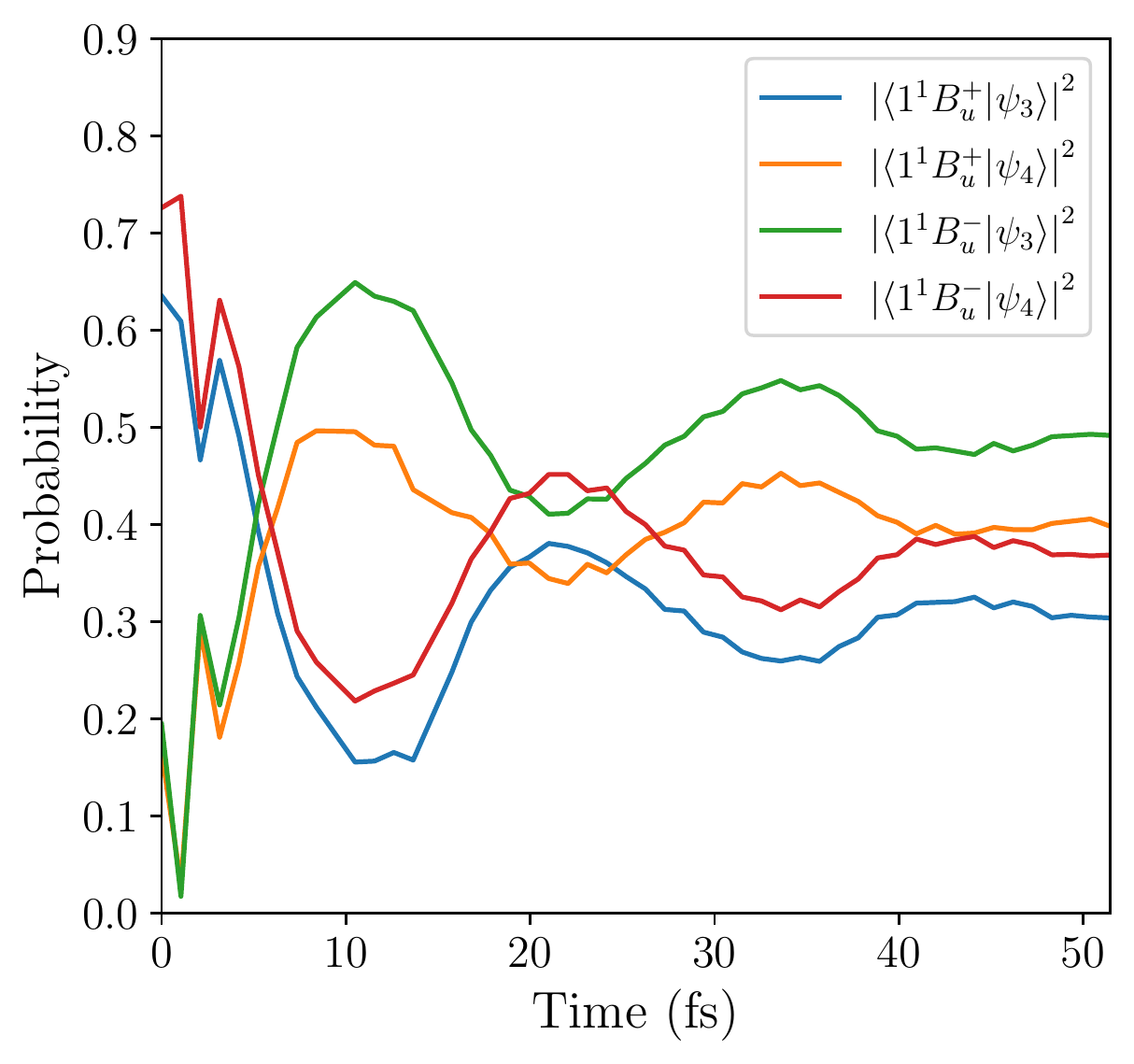}
\caption{Probabilities that the adiabatic states, $\ket{\psi_3}$ and $\ket{\psi_4}$, occupy the diabatic states, $\ketBup{1}$ and $\ketBum{1}$.
At time $t=0$, $\ket{\psi_3}$ is the lowest-energy photoexcited state, $S_2$.
}
\label{Fig:3}
\end{figure}

Fig.\ \ref{Fig:3} shows the probabilities that the adiabatic states occupy the diabatic states,  $\ketBup{1}$ and $\ketBum{1}$. Reflecting the crossover in the diabatic energies, at $t=0$ the lower adiabatic state, $\ket{\psi_3}$,   predominately occupies $\ketBup{1}$, while the upper adiabatic state, $\ket{\psi_4}$, predominately occupies $\ketBum{1}$.  At the avoided crossing  the adiabatic states are nearly equal admixtures of the diabatic states. These  probabilities then oscillate with a frequency approximately equal to the C-C bond vibration, before becoming damped after $\sim 40$ fs.
At this time the lower adiabatic state, $\ket{\psi_3}$, predominately occupies the triplet-pair state, $\ketBum{1}$.

As shown in Fig.\ \ref{Fig:2}, $\ket{\Psi(t)}$ is  $\sim 90\%$ composed of the adiabatic states $\ket{\psi_3}$ and $\ket{\psi_4}$. In addition, the adiabatic probabilities  ($|\psi_3|^2$ and $|\psi_4|^2$) and  energies ($E_3$ and $E_4$) become quasi-stationary after $\sim 30$ fs.
Thus, too a good approximation, we can adopt a two-level system and express $\ket{\Psi(t)}$ as the non-stationary state
\begin{equation}\label{}
  \ket{\Psi(t)} \approx \psi_3 \ket{\psi_3}\exp(-\textrm{i}E_3t/\hbar)  + \psi_4 \ket{\psi_4}\exp(-\textrm{i}E_4t/\hbar).
\end{equation}
Similarly, Fig.\ \ref{Fig:3}  shows that the adiabatic states are  $\sim 90\%$ composed of the diabatic states $\ketBup{1}$ and $\ketBum{1}$, i.e.,
\begin{equation}\label{}
  \ket{\psi_3} \approx a(t) \ketBup{1} + b(t) \ketBum{1}
\end{equation}
and
\begin{equation}\label{}
  \ket{\psi_4} \approx c(t) \ketBup{1} + d(t)\ketBum{1}.
\end{equation}
Thus, the probability that the system occupies the singlet triplet-pair  state is,
\begin{eqnarray}\label{Eq:9a}
\nonumber
 && P_{\Bum{1}}(t)  = \left| \langle \Bum{1}\ket{\Psi(t)}\right|^2 \\
                && \approx \left|b \psi_3 \right|^2 + \left|d \psi_4  \right|^2 +\left(b \psi_3  \right)^*\left(d \psi_4  \right)\cos(E_4-E_3)t/\hbar.
\end{eqnarray}
This  probability       
is illustrated in Fig.\ \ref{Fig:2} by the dashed-red curve. For $t \gtrsim 40$ fs it oscillates with a period $T = h/(E_4 - E_3) = 12$ fs, showing that Eq.\ (\ref{Eq:9a}) is justified.

In general, as well as causing oscillations in $ P_{\Bum{1}}(t)$, the quantum coherences between the adiabatic states cause time-dependent observables. In practice, however, interactions of the carotenoid chain (i.e., the system) with its surroundings will cause decoherence, and in  particular the oscillations in the probability that the system occupies the diabatic states $\ketBum{1}$ and $\ketBup{1}$ will be damped. These processes are not completely  modeled by our Ehrenfest approximation of the nuclear degrees of freedom, so we estimate the singlet triplet-pair yield by the `classical' component of Eq.\ (\ref{Eq:9a})  i.e.,
  $P_{\Bum{1}}^{\textrm{classical}} = \left|b \psi_3  \right|^2 + \left|d \psi_4  \right|^2$.
This yield is $\sim 45\%$ after $\sim 50$ fs.

As we have already noted, the relaxed $\Bum{1}$ state lies lower in energy than the relaxed quintet triplet-pair state, and as this quintet corresponds to a pair of spin-correlated but unbound triplets, we can conclude that potential intramolecular singlet fission via $\ketBum{1}$ is endothermic. As Fig.\ 1(b) in the SI indicates, however, intermolecular singlet fission via $\ketBum{1}$ on two carotenoid molecules of the same length is an exothermic process, because of an increase in (negative) nuclear reorganization energy and a decrease in (positive) confinement energy for single triplets on a chain.

Potential intramolecular singlet fission via $\ketAgm{3}$ is exothermic, because its excess kinetic energy overcomes the triplet binding energy. Indeed, as the polyene chain length increases, internal conversion  from $\ketBup{1}$ occurs to higher kinetic energy members of the `$2A_g$' family, meaning that for $N>24$ all internal conversion  is energetically favourable for intramolecular singlet fission.
In practice, since $\ketBum{1}$ and  $\ketAgm{3}$ are higher quasi-momentum counterparts of  $\ketAgm{2}$,  phonon-mediated internal conversion  from the former to the latter is possible. Or, a vibronically allowed internal conversion  from $S_2$ to $\ketAgm{2}$ might occur. We note, however, that singlet fission from  $\ketAgm{2}$ is expected to be endothermic for both intra and intermolecular processes.\footnote{This is a robust prediction over a wide range of model parameters, as indicated by Fig.\ 7.4 of Ref.\ \cite{Book}.}

In conclusion, our dynamical simulations of the photoexcited $S_2$ state of cartotenoids using a model of strongly correlated electrons show that internal conversion to a singlet triplet-pair state occurs adiabatically via an avoided crossing within 100 fs.
We further predict that  only intermolecular exothermic singlet fission is possible for shorter carotenoids,
but intramolecular exothermic singlet fission is possible for longer chains.
Although our theoretical predictions --  determined using the Chandross-Mazumdar\cite{Chandross97} parametrization of the PPP model -- are consistent with a wide range of experimental observations\cite{Rondonuwu03,Fujii04}, there does not exist a settled consensus about the relative orderings of the vertical energies of the singlet triplet-pair states and $S_2$, with some authors arguing that the  vertical energy of the $\Agm{2}$ state is higher than that of the $\Bup{1}$ state.\cite{Taffet19} Since the $\Bup{1}$ state is excitonic and thus is a fluctuating electric dipole,\cite{Barford11} its energy is strongly affected by the polarizability of its environment. This implies that in some environments the vertical $\Agm{2}$ energy might lie higher than the vertical $\Bup{1}$ energy and \textit{vice versa} for their relaxed energies, and therefore internal conversion is from the $\Bup{1}$ state to the $\Agm{2}$ state.

Future work will investigate internal conversion from $S_2$ to the $A_g$ sector and we will also examine the validity of the Ehrenfest approximation by quantizing the phonon degrees of freedom.


\begin{acknowledgement}
We thank Jenny Clark for helpful discussions.
D.M. receives financial support from the EPSRC Centre for Doctoral Training, Theory and Modelling in Chemical Sciences (Grant Ref.
EP/L015722/1), the Department of Chemistry, and Linacre College via the Carolyn and Franco Giantruco Scholarship.
D.J.V. received financial support from the EPSRC Centre for Doctoral Training, Theory and Modelling in Chemical Sciences (Grant Ref.
EP/L015722/1), the Department of Chemistry,  and Balliol College Oxford via the Foley-Béjar Scholarship.
M.M. received financial support from the UKRI (Grant Ref. EP/S002766/1).
\end{acknowledgement}


\begin{suppinfo}



\section{1.\ Parametrization of the Hamiltonian}\label{Se:1}

\subsection{1.1 Parametrization of $H_{\textrm{UV}}$}

As described in Section 2.2, efficient implementation of the time-dependent density matrix renormalization group method via the Trotter decomposition of the evolution operator requires that the Hamiltonian  be partitioned  into a sum of bond Hamiltonians. This implies that only on-site and nearest neighbor terms can be retained in the Hamiltonian. Thus, the purely  electronic component of the Hamiltonian is the extended-Hubbard (or UV) model. This model, defined by
\begin{equation}
\hat{H}_{\textrm{UV}} = -2\beta \sum_{n=1}^{N-1}  \hat{T}_n + U \sum_{n=1}^N  \big( \hat{N}_{n \uparrow} - \frac{1}{2} \big) \big( \hat{N}_{n \downarrow} - \frac{1}{2} \big)
+ \frac{1}{2} \sum_{n=1}^{N-1} V \big( \hat{N}_n -1 \big) \big( \hat{N}_{n+1} -1 \big),
\end{equation}
contains a nearest neighbor electron transfer term, $\beta$, and onsite and nearest neighbor Coulomb interactions, $U$ and $V$, respectively.
$\hat{T}_ n = \frac{1}{2}\sum_{\sigma} \left( c^{\dag}_{n , \sigma} c_{n + 1 , \sigma} + c^{\dag}_{n+ 1 , \sigma} c_{n  , \sigma} \right)$ is the bond order operator, $\hat{N}_n$ is the number operator and $N$ is the number of conjugated carbon-atoms ($N/2$ is the number of double bonds).

In recent work\cite{Valentine20} we studied the excited states of conjugated polyenes using the Pariser-Parr-Pople-Peierls (PPPP) model. The purely electronic component of this Hamiltonian, namely, the Pariser-Parr-Pople (PPP) model, contains nearest neighbor electron transfer terms and long range Coulomb interactions. We used the condensed-phase Chandross-Mazumdar\cite{Chandross97} parametrization of the PPP model.

In the current investigation we parametrized the UV model so that it replicates the predictions of the PPPP model with the Chandross-Mazumdar\cite{Chandross97} parametrization. We found that keeping $\beta= 2.4$ eV, $\omega_0 = \sqrt{K/m} = 2.15\times10^{14}$ s$^{-1}$  and $K = 46$ eV $\AA^{-2}$ as before, and choosing $U = 7.25$ eV, $V = 2.75$ eV  and $\alpha = 4.6$ eV $\AA^{-1}$ gives  quantitatively equivalent results, as shown by comparing Fig.\ 1\emph{} with Figures 2 and 3 of Ref.\cite{Valentine20}.

\begin{figure}[h!]\label{Fig:SI1}
    \centering
    \includegraphics[width=1.65\myfigwidth]{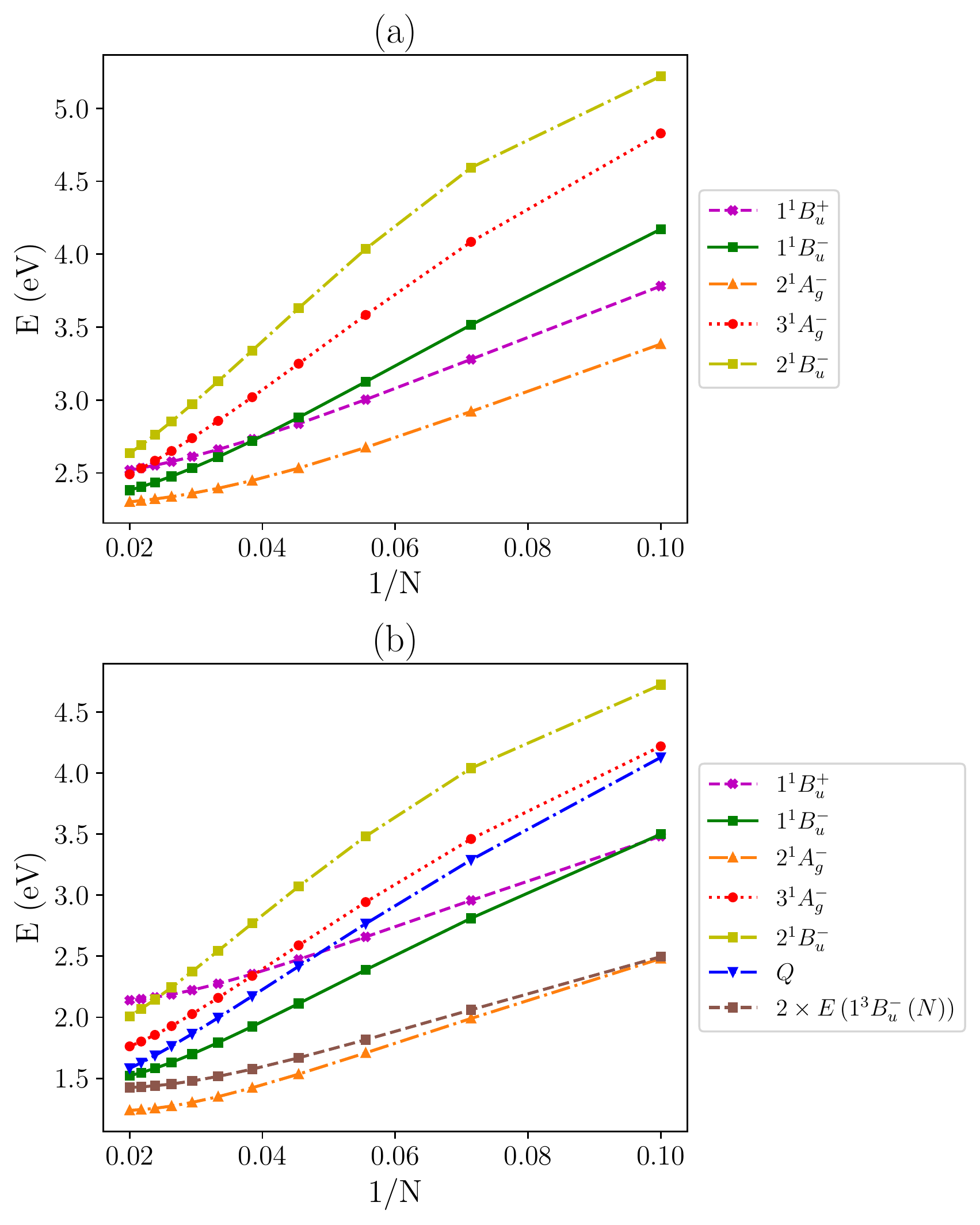}
    \caption{\small{Vertical (a) and relaxed (b) diabatic singlet excitation energies of the UV-Peierls model. $N$ is the number of conjugated carbon atoms and $N/2$ is the number of double bonds.
    These results indicate that internal conversion from $\ketBup{1}$ to $\ketBum{1}$ is energetically possible for  $12 \leq N \leq 24$, while  internal conversion from $\ketBup{1}$ to $\ketAgm{3}$ is energetically possible for  $26 \leq N \leq 42$.
    Also shown in (b) is the quintet energy and twice the lowest triplet energy, implying that  (i) singlet fission from $\ketAgm{2}$ is endothermic for both intra and intermolecular processes, (ii) singlet fission from $\ketBum{1}$ is endothermic for intramolecular and exothermic for intermolecular processes, and (iii) singlet fission from $\ketAgm{3}$ is exothermic for both intra and intermolecular processes.}}
\end{figure}

\subsection{1.2 Parametrization of $H_{\textrm{SB}}$}

The substituents on carotenoid molecules causes the particle-hole symmetry of polyene chains to be broken, which implies that the $\pi$-electron Mulliken charge densities\cite{Mulliken55} are not equal to unity on every carbon-atom.
The  Mulliken charge densities for zeaxanthin were calculated using the ORCA programme package\cite{Neese12,Neese17}. To that end a geometry optimisation was performed using density functional theory (DFT) with a B3LYP functional\cite{Stephens94} and a def2-TZVP basis set\cite{Weigend05,Weigend06} before calculating the electron densities.  The tabulated values, shown in Table I, are the $\pi$-electron charges associated with each carbon atom along the conjugated backbone, numbered 1-22 in Fig.\ 2.

We ensure that our $\pi$-electron model replicates the DFT-calculated charges (subject to  charge neutrality) by supplementing $\hat{H}_{\textrm{UV}}$ with
\begin{equation}
	\hat{H}_{\textrm{SB}} =  \sum_{n=1}^{N} \epsilon_n (\hat{N}_n-1),
\end{equation}
where the potential energy terms are parametrized to replicate the DFT calculations. These are listed in Table I.

\begin{figure*}[h!]\label{Fig:SI2}
    \centering
    \includegraphics[width=1.6\myfigwidth]{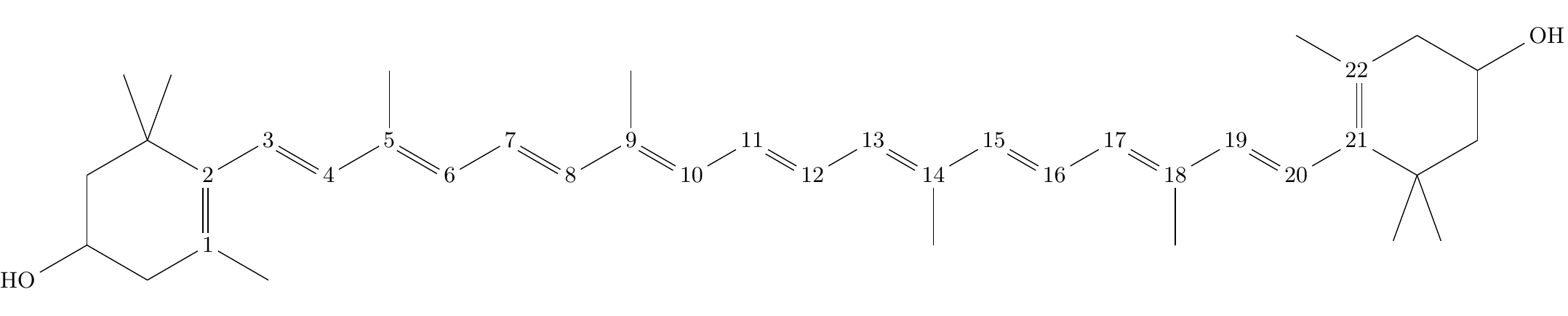}
    \caption{Structural formula of zeaxanthin.}
\end{figure*}

\begin{table}
\small\centering
{\renewcommand{\arraystretch}{1.2}
\begin{tabular}{|p{0.20\linewidth}|p{0.37\linewidth}|p{0.20\linewidth}|}
\hline
C-site, $n$  & Mulliken charges ($q$) &$\epsilon_n$ (eV)  \\
\hline
1 & 1.34 & -3.6   \\
2 & 0.94 & -1.0  \\
3 & 1.04 & 0.0  \\
4 & 0.97 & 0.0   \\
5 & 1.22 & -1.6  \\
6 & 0.98 & -0.6 \\
7 & 1.05 & 0.0   \\
8 & 0.98 & 0.0  \\
9 & 1.22 & -1.6  \\
10 & 0.98 & -0.6   \\
11 & 1.03 & 0.0  \\
\hline
\end{tabular}}
\caption{The $\pi$-electron Mulliken charges from the \emph{ab initio} DFT calculation  and the parameters for $\hat{H}_{\textrm{SB}}$.
In order to maintain $\pi$-electron charge neutrality, each \emph{ab initio} charge was reduced by $0.07 q$.
$\epsilon_n = + \epsilon_{N+1-n}$ for even $\hat{H}_{\textrm{SB}}$, implying a $\ketBup{1}$ to $\ketBum{1}$ crossover.
The carbon-sites are labelled in Fig.\ 2.}
\label{ta:1}
\end{table}

\section{2.\ Computational Methods}\label{Se:2}

\subsection{2.1 Determining the ground state}

The dimerized ground state is determined by solving
\begin{equation}
	\hat{H} = \hat{H}_{\textrm{UV}} + \hat{H}_{\textrm{e-n}}+ \hat{H}_{\textrm{SB}} + \frac{K}{2} \sum_{n=1}^{N-1} \left( u_{n+1} - u_n \right)^2 \label{D1}
\end{equation}
for fixed nuclear displacements, $u_n$, using the static density matrix renomalization group method\cite{White92}. The equilibrium displacements are found by iterative application of Eq.\ (\ref{Eq:9}) with  $f_n = 0$.\cite{Barford01}

\subsection{2.2 Solution of the time-dependent Schr\"odinger equation}\label{Se:2.2}

The dynamics of the evolving system is fully determined by solving the  time-dependent Schr\"odinger equation
\begin{equation}
\textrm{i} \hbar\frac{\textrm{d}\ket{\Psi(t)}}{\textrm{d} t} = \hat{H}\ket{\Psi(t)} .	\label{D2}
\end{equation}
Eq. (\ref{D2}) has the formal solution
\begin{equation}
	\ket{\Psi(t+\delta t)} = e^{-\textrm{i}\hat{H} \delta t/\hbar}\ket{\Psi(t)},
\end{equation}
where the initial condition is $\ket{\Psi(t=0)} = \hat{\mu} \ket{\textrm{GS}}$ and $\ket{\textrm{GS}}$ is the ground state solution of Eq.\ (\ref{D1}).

$\ket{\Psi(t)}$ is determined using the adaptive time-dependent density matrix renormalisation group (t-DMRG) method, where the truncated Hilbert space is updated such that loss of information about the system is minimized.\cite{White04,Daley2004} Since the full Hamiltonian only contains on-site and nearest neighbor terms, it can be written as a sum of bond Hamiltonians, i.e.,
\begin{equation}
\hat{H} = \hat{H}_{1,2} + \hat{H}_{2,3} + \hat{H}_{3,4} +\cdots + \hat{H}_{n,n+1} + \cdots + \hat{H}_{N-1,N},
\end{equation}
where $\hat{H}_{n,n+1}$ acts on the $n$th.\ bond.
A Suzuki-Trotter decomposition is invoked for the propagator, i.e.,
\begin{eqnarray}\nonumber
	e^{-\textrm{i}\hat{H} \delta t/\hbar} \approx e^{-\textrm{i}\hat{H}_{1,2} \delta t /2\hbar} e^{-\textrm{i}\hat{H}_{2,3} \delta t /2\hbar} \cdots e^{-\textrm{i}\hat{H}_{N-1,N} \delta t /2\hbar} e^{-\textrm{i}\hat{H}_{N-1,N} \delta t /2\hbar} \cdots e^{-\textrm{i}\hat{H}_{2,3} \delta t /2\hbar} e^{-\textrm{i}\hat{H}_{1,2} \delta t /2\hbar}\\
 + O(\delta t^3)
\end{eqnarray}
The propagator is applied $t/\delta t$ times on $\ket{\Psi(t=0)}$ to find $\ket{\Psi(t)}$ with $\delta t = 3 \times 10^{-4} \textrm{ fs}$.

\subsection{2.3 Ehrenfest equations of motion}\label{Se:2.3}

The nuclear degrees of freedom, defined by Eqs. (2) and (3) of the main paper, 
are treated classically via the Ehrenfest equations of motion. The force on atom $n$ is
\begin{equation}
f_{n} = - \frac{\textrm{d} \langle \Psi | \hat{H} \ket{\Psi}}{\textrm{d} u_{n}}
	= - \left \langle \frac{\textrm{d} \hat{H}}{\textrm{d} u_{n}} \right \rangle
	= - \left \langle \frac{\textrm{d} \hat{H}_{\textrm{e-n}}}{\textrm{d} u_{n}} \right \rangle
	- \left \langle \frac{\textrm{d} \hat{H}_{\textrm{nuclear}}}{\textrm{d} u_{n}} \right \rangle,
\end{equation}
i.e.,
\begin{equation}\label{Eq:9}
f_{n} 	= 2 \alpha  \left (	\langle \hat{T}_n	 \rangle - \langle \hat{T}_{n-1} \rangle \right	)
	- K  \left  (2 u_{n} - u_{n+1} - u_{n-1}			\right	).
\end{equation}
The nuclei obey the coupled equations of motion
\begin{equation}
	\frac{\textrm{d} u_{n}(t)}{\textrm{d} t} = \frac{p_{n}(t)}{m}
\end{equation}
and
\begin{equation}
	\frac{\textrm{d} p_{n}(t)}{\textrm{d} t} = f_{n}(t) - \gamma p_{n} (t),
\end{equation}
where a phenomenological linear damping term $\gamma p_{n}$ is introduced to cause relaxation of the nuclei. The equations of
motion are propagated using the damped Velocity Verlet scheme (derived in Appendix 5.A of Ref\cite{Valentine2020b}), i.e., 
\begin{equation}
u_{n}(t+\Delta t)=u_{n}(t)+\frac{p_{n}(t)}{m} \Delta t+\frac{1}{2} \frac{(f_{n}(t) - \gamma p_{n}(t))}{m} \Delta t^{2}
\end{equation}
and
\begin{equation}
	p_{n}(t+\Delta t)=\frac{1}{(1+\gamma\Delta t  / 2)}\left((1-\gamma\Delta t  / 2)p_{n}(t)
	+\frac{\Delta t}{2 }\left(f_{n}(t+\Delta t)+f_{n}(t)\right)\right).
\end{equation}
We use $\gamma = 1.52 \times 10^{14} \textrm{ s}^{-1}$, corresponding to $\gamma = {\omega_D}/{2}
= {\omega}/{\sqrt{2}}$ such that critical damping is not reached. The integration time step is set to $\Delta t = 0.003 \textrm{ fs}$ such that $\Delta t ^\prime = \omega_D \Delta t = 9 \times 10^{-4} \ll 1$.

\subsection{2.4 Accuracy and convergence tests}

Due to the variational nature of the DMRG algorithm, its accuracy can be systematically improved  by  increasing the Hilbert space. Convergence of the closely related Pariser-Parr-Pople-Peierls model has been extensively studied.\cite{Bursill1999,Barford2002c,Bursill2002,Bursill2009}

In general, in t-DMRG the number of states required to represent a time-evolving state vector accurately is given by
\begin{equation}
m = 2^S,	\label{D3}
\end{equation}
where $S$ is the von Neumann entanglement entropy. For our simulations, the maximum entropy reached by a block is
$S_{\textrm{max}}=2.46$. We typically retain over 400 states per block, which is much more than the number of states required
by Eq.\ (\ref{D3}).

\section{3.\ Dynamics at the Avoided Crossing}

Fig.\ 3 of the main paper shows that $\ket{\Psi(t)}$ is  $\sim 90\%$ composed of the adiabatic states $\ket{\psi_3}$ and $\ket{\psi_4}$.
In addition, the adiabatic probabilities  ($\psi_3^2$ and $\psi_4^2$) and  energies ($E_3$ and $E_4$) become quasi-stationary after $\sim 30$ fs.
Thus, too a good approximation, we can adopt a two-level system and express $\ket{\Psi(t)}$ as the non-stationary state
\begin{equation}\label{}
  \ket{\Psi(t)} \approx \psi_3 \ket{\psi_3}\exp(-\textrm{i}E_3t/\hbar)  + \psi_4 \ket{\psi_4}\exp(-\textrm{i}E_4t/\hbar).
\end{equation}
Similarly, Fig.\ 4 of the main paper shows that the adiabatic states are  $\sim 90\%$ composed of the diabatic states $\ketBup{1}$ and $\ketBum{1}$, i.e.,
\begin{equation}\label{}
  \ket{\psi_3} \approx a(t) \ketBup{1} + b(t) \ketBum{1}
\end{equation}
and
\begin{equation}\label{}
  \ket{\psi_4} \approx c(t) \ketBup{1} + d(t)\ketBum{1}.
\end{equation}
In a two-level system, $|a|^2 = |d|^2$ and $|c|^2 = |b|^2$. As shown by Fig.\ 4 of the main paper, however, these conditions are not exactly satisfied, so our system is not precisely a two-level system.

The evolution of $\ket{\Psi(t)}$ is determined by
\begin{equation}\label{}
  \hat{H} = \hat{H}_{\textrm{UV}} + \hat{H}_{\textrm{e-n}} + \hat{H}_{\textrm{SB}}.
\end{equation}
$(\hat{H}_{\textrm{UV}} + \hat{H}_{\textrm{e-n}})$ is even under both a $C_{2h}$ operation and a particle-hole transformation, whereas  $\hat{H}_{\textrm{SB}}$ is odd under a particle-hole  transformation and (in our model)  even  under a $C_{2h}$ operation. Thus, in the $2\times 2$ basis of the diabatic states of $\ketBup{1}$ and  $\ketBum{1}$ the Hamiltonian is block-diagonalized by $(\hat{H}_{\textrm{UV}} + \hat{H}_{\textrm{e-n}})$ and $\hat{H}_{\textrm{SB}}$.

Therefore, the Hamiltonian matrix is
\begin{equation}\label{}
  \begin{pmatrix}
E_{\Bup{1}} & V \\
V^* & E_{\Bum{1}}
\end{pmatrix},
\end{equation}
where $E_X = \langle X|\hat{H}_{\textrm{UV}} + \hat{H}_{\textrm{e-n}}|X\rangle$ and $V = \langle \Bup{1}|\hat{H}_{\textrm{SB}}|\Bum{1}\rangle$.
The instantaneous eigenvalues are
\begin{equation}\label{}
  E_{\pm} = \frac{E_{\Bup{1}}+E_{\Bum{1}}}{2} \pm \left( \left(\frac{E_{\Bup{1}}-E_{\Bum{1}}}{2}\right)^2 + |V|^2\right)^{1/2},
\end{equation}
where we associate $E_3$ with $E_-$ and $E_4$ with $E_+$.

The coupling matrix element, $V$, remains finite during the evolution and so at the energy crossing of the diabatics (i.e., when $E_{\Bup{1}}$=$E_{\Bum{1}}$) there is an avoided crossing with a gap $\Delta E = (E_+-E_-) = 2|V| = 0.23 $ eV, as can be seen in Fig.\ 2 of the main paper.

\end{suppinfo}

\bibliography{references}

\providecommand{\latin}[1]{#1}
\makeatletter
\providecommand{\doi}
  {\begingroup\let\do\@makeother\dospecials
  \catcode`\{=1 \catcode`\}=2 \doi@aux}
\providecommand{\doi@aux}[1]{\endgroup\texttt{#1}}
\makeatother
\providecommand*\mcitethebibliography{\thebibliography}
\csname @ifundefined\endcsname{endmcitethebibliography}
  {\let\endmcitethebibliography\endthebibliography}{}
\begin{mcitethebibliography}{38}
\providecommand*\natexlab[1]{#1}
\providecommand*\mciteSetBstSublistMode[1]{}
\providecommand*\mciteSetBstMaxWidthForm[2]{}
\providecommand*\mciteBstWouldAddEndPuncttrue
  {\def\EndOfBibitem{\unskip.}}
\providecommand*\mciteBstWouldAddEndPunctfalse
  {\let\EndOfBibitem\relax}
\providecommand*\mciteSetBstMidEndSepPunct[3]{}
\providecommand*\mciteSetBstSublistLabelBeginEnd[3]{}
\providecommand*\EndOfBibitem{}
\mciteSetBstSublistMode{f}
\mciteSetBstMaxWidthForm{subitem}{(\alph{mcitesubitemcount})}
\mciteSetBstSublistLabelBeginEnd
  {\mcitemaxwidthsubitemform\space}
  {\relax}
  {\relax}

\bibitem[Hudson and Kohler(1972)Hudson, and Kohler]{Hudson72}
Hudson,~B.~S.; Kohler,~B.~E. Low-Lying Weak Transition in Polyene Alpha,
  Omega-Diphenyloctatetraene. \emph{Chemical Physics Letters} \textbf{1972},
  \emph{14}, 299\relax
\mciteBstWouldAddEndPuncttrue
\mciteSetBstMidEndSepPunct{\mcitedefaultmidpunct}
{\mcitedefaultendpunct}{\mcitedefaultseppunct}\relax
\EndOfBibitem
\bibitem[Schulten and Karplus(1972)Schulten, and Karplus]{Schulten72}
Schulten,~K.; Karplus,~M. Origin of a Low-Lying Forbidden Transition in
  Polyenes and Related Molecules. \emph{Chemical Physics Letters}
  \textbf{1972}, \emph{14}, 305\relax
\mciteBstWouldAddEndPuncttrue
\mciteSetBstMidEndSepPunct{\mcitedefaultmidpunct}
{\mcitedefaultendpunct}{\mcitedefaultseppunct}\relax
\EndOfBibitem
\bibitem[Hayden and Mele(1986)Hayden, and Mele]{Hayden86}
Hayden,~G.~W.; Mele,~E.~J. Correlation-Effects and Excited-States in Conjugated
  Polymers. \emph{Physical Review B} \textbf{1986}, \emph{34}, 5484--5497\relax
\mciteBstWouldAddEndPuncttrue
\mciteSetBstMidEndSepPunct{\mcitedefaultmidpunct}
{\mcitedefaultendpunct}{\mcitedefaultseppunct}\relax
\EndOfBibitem
\bibitem[Tavan and Schulten(1987)Tavan, and Schulten]{Tavan87}
Tavan,~P.; Schulten,~K. Electronic Excitations in Finite and Infinite Polyenes.
  \emph{Physical Review B} \textbf{1987}, \emph{36}, 4337--4358\relax
\mciteBstWouldAddEndPuncttrue
\mciteSetBstMidEndSepPunct{\mcitedefaultmidpunct}
{\mcitedefaultendpunct}{\mcitedefaultseppunct}\relax
\EndOfBibitem
\bibitem[Barford(2013)]{Book}
Barford,~W. \emph{Electronic and optical properties of conjugated polymers},
  2nd ed.; Oxford University Press: Oxford, 2013\relax
\mciteBstWouldAddEndPuncttrue
\mciteSetBstMidEndSepPunct{\mcitedefaultmidpunct}
{\mcitedefaultendpunct}{\mcitedefaultseppunct}\relax
\EndOfBibitem
\bibitem[Kraabel \latin{et~al.}(1998)Kraabel, Hulin, Aslangul,
  Lapersonne-Meyer, and Schott]{Kraabel98}
Kraabel,~B.; Hulin,~D.; Aslangul,~C.; Lapersonne-Meyer,~C.; Schott,~M. Triplet
  exciton generation, transport and relaxation in isolated polydiacetylene
  chains: subpicosecond pump-probe experiments. \emph{Chemical Physics}
  \textbf{1998}, \emph{227}, 83--98\relax
\mciteBstWouldAddEndPuncttrue
\mciteSetBstMidEndSepPunct{\mcitedefaultmidpunct}
{\mcitedefaultendpunct}{\mcitedefaultseppunct}\relax
\EndOfBibitem
\bibitem[Lanzani \latin{et~al.}(1999)Lanzani, Stagira, Cerullo, De~Silvestri,
  Comoretto, Moggio, Cuniberti, Musso, and Dellepiane]{Lanzani99}
Lanzani,~G.; Stagira,~S.; Cerullo,~G.; De~Silvestri,~S.; Comoretto,~D.;
  Moggio,~I.; Cuniberti,~C.; Musso,~G.~F.; Dellepiane,~G. Triplet exciton
  generation and decay in a red polydiacetylene studied by femtosecond
  spectroscopy. \emph{Chemical Physics Letters} \textbf{1999}, \emph{313},
  525--532\relax
\mciteBstWouldAddEndPuncttrue
\mciteSetBstMidEndSepPunct{\mcitedefaultmidpunct}
{\mcitedefaultendpunct}{\mcitedefaultseppunct}\relax
\EndOfBibitem
\bibitem[Musser \latin{et~al.}(2013)Musser, Al-Hashimi, Maiuri, Brida, Heeney,
  Cerullo, Friend, and Clark]{Musser13}
Musser,~A.~J.; Al-Hashimi,~M.; Maiuri,~M.; Brida,~D.; Heeney,~M.; Cerullo,~G.;
  Friend,~R.~H.; Clark,~J. Activated Singlet Exciton Fission in a
  Semiconducting Polymer. \emph{Journal of the American Chemical Society}
  \textbf{2013}, \emph{135}, 12747--12754\relax
\mciteBstWouldAddEndPuncttrue
\mciteSetBstMidEndSepPunct{\mcitedefaultmidpunct}
{\mcitedefaultendpunct}{\mcitedefaultseppunct}\relax
\EndOfBibitem
\bibitem[Kasai \latin{et~al.}(2015)Kasai, Tamai, Ohkita, Benten, and
  Ito]{Kasai15}
Kasai,~Y.; Tamai,~Y.; Ohkita,~H.; Benten,~H.; Ito,~S. Ultrafast Singlet Fission
  in a Push-Pull Low-Bandgap Polymer Film. \emph{Journal of the American
  Chemical Society} \textbf{2015}, \emph{137}, 15980--15983\relax
\mciteBstWouldAddEndPuncttrue
\mciteSetBstMidEndSepPunct{\mcitedefaultmidpunct}
{\mcitedefaultendpunct}{\mcitedefaultseppunct}\relax
\EndOfBibitem
\bibitem[Busby \latin{et~al.}(2015)Busby, Xia, Wu, Low, Song, Miller, Zhu,
  Campos, and Sfeir]{Busby15}
Busby,~E.; Xia,~J.~L.; Wu,~Q.; Low,~J.~Z.; Song,~R.; Miller,~J.~R.; Zhu,~X.~Y.;
  Campos,~L.~M.; Sfeir,~M.~Y. A design strategy for intramolecular singlet
  fission mediated by charge-transfer states in donor-acceptor organic
  materials. \emph{Nature Materials} \textbf{2015}, \emph{14}, 426--433\relax
\mciteBstWouldAddEndPuncttrue
\mciteSetBstMidEndSepPunct{\mcitedefaultmidpunct}
{\mcitedefaultendpunct}{\mcitedefaultseppunct}\relax
\EndOfBibitem
\bibitem[Huynh \latin{et~al.}(2017)Huynh, Basel, Ehrenfreund, Li, Yang,
  Mazumdar, and Vardeny]{Huynh17}
Huynh,~U. N.~V.; Basel,~T.~P.; Ehrenfreund,~E.; Li,~G.; Yang,~Y.; Mazumdar,~S.;
  Vardeny,~Z.~V. Transient Magnetophotoinduced Absorption Studies of
  Photoexcitations in $\pi$-Conjugated Donor-Acceptor Copolymers.
  \emph{Physical Review Letters} \textbf{2017}, \emph{119}\relax
\mciteBstWouldAddEndPuncttrue
\mciteSetBstMidEndSepPunct{\mcitedefaultmidpunct}
{\mcitedefaultendpunct}{\mcitedefaultseppunct}\relax
\EndOfBibitem
\bibitem[Musser and Clark(2019)Musser, and Clark]{Musser19}
Musser,~A.~J.; Clark,~J. Triplet-Pair States in Organic Semiconductors.
  \emph{Annual Review of Physical Chemistry, Vol 70} \textbf{2019}, \emph{70},
  323--351\relax
\mciteBstWouldAddEndPuncttrue
\mciteSetBstMidEndSepPunct{\mcitedefaultmidpunct}
{\mcitedefaultendpunct}{\mcitedefaultseppunct}\relax
\EndOfBibitem
\bibitem[Sanders \latin{et~al.}(2019)Sanders, Pun, Parenti, Kumarasamy, Yablon,
  Sfeir, and Compos]{Sanders19}
Sanders,~S.~N.; Pun,~A.~B.; Parenti,~K.~R.; Kumarasamy,~E.; Yablon,~L.~M.;
  Sfeir,~M.~Y.; Compos,~L.~M. Understanding the Bound Triplet-Pair State in
  Singlet Fission. \emph{Chem} \textbf{2019}, \emph{5}, 1988--2005\relax
\mciteBstWouldAddEndPuncttrue
\mciteSetBstMidEndSepPunct{\mcitedefaultmidpunct}
{\mcitedefaultendpunct}{\mcitedefaultseppunct}\relax
\EndOfBibitem
\bibitem[Valentine \latin{et~al.}(2020)Valentine, Manawadu, and
  Barford]{Valentine20}
Valentine,~D.~J.; Manawadu,~D.; Barford,~W. Higher-energy triplet-pair states
  in polyenes and their role in intramolecular singlet fission. \emph{Physical
  Review B} \textbf{2020}, \emph{102}\relax
\mciteBstWouldAddEndPuncttrue
\mciteSetBstMidEndSepPunct{\mcitedefaultmidpunct}
{\mcitedefaultendpunct}{\mcitedefaultseppunct}\relax
\EndOfBibitem
\bibitem[Taffet \latin{et~al.}(2020)Taffet, Beljonne, and Scholes]{Taffet20}
Taffet,~E.~J.; Beljonne,~D.; Scholes,~G.~D. Overlap-Driven Splitting of Triplet
  Pairs in Singlet Fission. \emph{Journal of the American Chemical Society}
  \textbf{2020}, \emph{142}, 20040--20047\relax
\mciteBstWouldAddEndPuncttrue
\mciteSetBstMidEndSepPunct{\mcitedefaultmidpunct}
{\mcitedefaultendpunct}{\mcitedefaultseppunct}\relax
\EndOfBibitem
\bibitem[Chandross and Mazumdar(1997)Chandross, and Mazumdar]{Chandross97}
Chandross,~M.; Mazumdar,~S. Coulomb interactions and linear, nonlinear, and
  triplet absorption in poly(para-phenylenevinylene). \emph{Physical Review B}
  \textbf{1997}, \emph{55}, 1497--1504\relax
\mciteBstWouldAddEndPuncttrue
\mciteSetBstMidEndSepPunct{\mcitedefaultmidpunct}
{\mcitedefaultendpunct}{\mcitedefaultseppunct}\relax
\EndOfBibitem
\bibitem[Rondonuwu \latin{et~al.}(2003)Rondonuwu, Watanabe, Fujii, and
  Koyama]{Rondonuwu03}
Rondonuwu,~F.~S.; Watanabe,~Y.; Fujii,~R.; Koyama,~Y. A first detection of
  singlet to triplet conversion from the $1^1B_u^-$ to the $1^3A_g$ state and
  triplet internal conversion from the $1^3A_g$ to the $1^3B_u$ state in
  carotenoids: dependence on the conjugation length. \emph{Chemical Physics
  Letters} \textbf{2003}, \emph{376}, 292--301\relax
\mciteBstWouldAddEndPuncttrue
\mciteSetBstMidEndSepPunct{\mcitedefaultmidpunct}
{\mcitedefaultendpunct}{\mcitedefaultseppunct}\relax
\EndOfBibitem
\bibitem[White and Feiguin(2004)White, and Feiguin]{White04}
White,~S.~R.; Feiguin,~A.~E. Real-time evolution using the density matrix
  renormalization group. \emph{Physical Review Letters} \textbf{2004},
  \emph{93}\relax
\mciteBstWouldAddEndPuncttrue
\mciteSetBstMidEndSepPunct{\mcitedefaultmidpunct}
{\mcitedefaultendpunct}{\mcitedefaultseppunct}\relax
\EndOfBibitem
\bibitem[Barford \latin{et~al.}(2001)Barford, Bursill, and
  Lavrentiev]{Barford01}
Barford,~W.; Bursill,~R.~J.; Lavrentiev,~M.~Y. Density-matrix
  renormalization-group calculations of excited states of linear polyenes.
  \emph{Physical Review B} \textbf{2001}, \emph{63}\relax
\mciteBstWouldAddEndPuncttrue
\mciteSetBstMidEndSepPunct{\mcitedefaultmidpunct}
{\mcitedefaultendpunct}{\mcitedefaultseppunct}\relax
\EndOfBibitem
\bibitem[White(1992)]{White92}
White,~S.~R. Density-Matrix Formulation for Quantum Renormalization-Groups.
  \emph{Physical Review Letters} \textbf{1992}, \emph{69}, 2863--2866\relax
\mciteBstWouldAddEndPuncttrue
\mciteSetBstMidEndSepPunct{\mcitedefaultmidpunct}
{\mcitedefaultendpunct}{\mcitedefaultseppunct}\relax
\EndOfBibitem
\bibitem[Horsfield \latin{et~al.}(2006)Horsfield, Bowler, Ness, Sanchez,
  Todorov, and Fisher]{Horsfield06}
Horsfield,~A.~P.; Bowler,~D.~R.; Ness,~H.; Sanchez,~C.~G.; Todorov,~T.~N.;
  Fisher,~A.~J. The transfer of energy between electrons and ions in solids.
  \emph{Reports on Progress in Physics} \textbf{2006}, \emph{69},
  1195--1234\relax
\mciteBstWouldAddEndPuncttrue
\mciteSetBstMidEndSepPunct{\mcitedefaultmidpunct}
{\mcitedefaultendpunct}{\mcitedefaultseppunct}\relax
\EndOfBibitem
\bibitem[Tully(2012)]{Tully12}
Tully,~J.~C. Perspective: Nonadiabatic dynamics theory. \emph{Journal of
  Chemical Physics} \textbf{2012}, \emph{137}\relax
\mciteBstWouldAddEndPuncttrue
\mciteSetBstMidEndSepPunct{\mcitedefaultmidpunct}
{\mcitedefaultendpunct}{\mcitedefaultseppunct}\relax
\EndOfBibitem
\bibitem[Fujii \latin{et~al.}(2004)Fujii, Fujino, Inaba, Nagae, and
  Koyama]{Fujii04}
Fujii,~R.; Fujino,~T.; Inaba,~T.; Nagae,~H.; Koyama,~Y. Internal conversion of
  $1B_u^+ \rightarrow 1B_u^- \rightarrow 2A_g^-$ and fluorescence from the
  $1B_u^-$ state in all-trans-neurosporene as probed by up-conversion
  spectroscopy. \emph{Chemical Physics Letters} \textbf{2004}, \emph{384},
  9--15\relax
\mciteBstWouldAddEndPuncttrue
\mciteSetBstMidEndSepPunct{\mcitedefaultmidpunct}
{\mcitedefaultendpunct}{\mcitedefaultseppunct}\relax
\EndOfBibitem
\bibitem[Taffet \latin{et~al.}(2019)Taffet, Lee, Toa, Pace, Rumbles, Southall,
  Cogdell, and Scholes]{Taffet19}
Taffet,~E.~J.; Lee,~B.~G.; Toa,~Z. S.~D.; Pace,~N.; Rumbles,~G.; Southall,~J.;
  Cogdell,~R.~J.; Scholes,~G.~D. Carotenoid Nuclear Reorganization and
  Interplay of Bright and Dark Excited States. \emph{Journal of Physical
  Chemistry B} \textbf{2019}, \emph{123}, 8628--8643\relax
\mciteBstWouldAddEndPuncttrue
\mciteSetBstMidEndSepPunct{\mcitedefaultmidpunct}
{\mcitedefaultendpunct}{\mcitedefaultseppunct}\relax
\EndOfBibitem
\bibitem[Barford \latin{et~al.}(2011)Barford, Paiboonvorachat, and
  Yaron]{Barford11}
Barford,~W.; Paiboonvorachat,~N.; Yaron,~D. Second-order dispersion
  interactions in $\pi$-conjugated polymers. \emph{Journal of Chemical Physics}
  \textbf{2011}, \emph{134}\relax
\mciteBstWouldAddEndPuncttrue
\mciteSetBstMidEndSepPunct{\mcitedefaultmidpunct}
{\mcitedefaultendpunct}{\mcitedefaultseppunct}\relax
\EndOfBibitem
\bibitem[Mulliken(1955)]{Mulliken55}
Mulliken,~R.~S. Electronic Population Analysis on LCAO–MO Molecular Wave
  Functions. I. \emph{Journal of Chemical Physics} \textbf{1955}, \emph{23},
  1833\relax
\mciteBstWouldAddEndPuncttrue
\mciteSetBstMidEndSepPunct{\mcitedefaultmidpunct}
{\mcitedefaultendpunct}{\mcitedefaultseppunct}\relax
\EndOfBibitem
\bibitem[Neese(2012)]{Neese12}
Neese,~F. The ORCA program system. \emph{Wiley Interdisciplinary Reviews:
  Computational Molecular Science} \textbf{2012}, \emph{2}, 73\relax
\mciteBstWouldAddEndPuncttrue
\mciteSetBstMidEndSepPunct{\mcitedefaultmidpunct}
{\mcitedefaultendpunct}{\mcitedefaultseppunct}\relax
\EndOfBibitem
\bibitem[Neese(2017)]{Neese17}
Neese,~F. Software update: the ORCA program system, version 4.0. \emph{Wiley
  Interdisciplinary Reviews: Computational Molecular Science} \textbf{2017},
  \emph{8}, e1327\relax
\mciteBstWouldAddEndPuncttrue
\mciteSetBstMidEndSepPunct{\mcitedefaultmidpunct}
{\mcitedefaultendpunct}{\mcitedefaultseppunct}\relax
\EndOfBibitem
\bibitem[Stephens \latin{et~al.}(1994)Stephens, Devlin, Chabalowski, and
  Frisch]{Stephens94}
Stephens,~P.~J.; Devlin,~F.~J.; Chabalowski,~C.~F.; Frisch,~M.~J. Ab Initio
  Calculation of Vibrational Absorption and Circular Dichroism Spectra Using
  Density Functional Force Fields. \emph{J. Phys. Chem.} \textbf{1994},
  \emph{98}, 11623\relax
\mciteBstWouldAddEndPuncttrue
\mciteSetBstMidEndSepPunct{\mcitedefaultmidpunct}
{\mcitedefaultendpunct}{\mcitedefaultseppunct}\relax
\EndOfBibitem
\bibitem[Weigend and Ahlrichs(2005)Weigend, and Ahlrichs]{Weigend05}
Weigend,~F.; Ahlrichs,~R. Balanced basis sets of split valence, triple zeta
  valence and quadruple zeta valence quality for H to Rn: Design and assessment
  of accuracy. \emph{Phys. Chem. Chem. Phys.} \textbf{2005}, \emph{7},
  3297\relax
\mciteBstWouldAddEndPuncttrue
\mciteSetBstMidEndSepPunct{\mcitedefaultmidpunct}
{\mcitedefaultendpunct}{\mcitedefaultseppunct}\relax
\EndOfBibitem
\bibitem[Weigend and Ahlrichs(2006)Weigend, and Ahlrichs]{Weigend06}
Weigend,~F.; Ahlrichs,~R. Accurate Coulomb-fitting basis sets for H to Rn.
  \emph{Phys. Chem. Chem. Phys.} \textbf{2006}, \emph{8}, 1057\relax
\mciteBstWouldAddEndPuncttrue
\mciteSetBstMidEndSepPunct{\mcitedefaultmidpunct}
{\mcitedefaultendpunct}{\mcitedefaultseppunct}\relax
\EndOfBibitem
\bibitem[Daley \latin{et~al.}(2004)Daley, Kollath, Schollw{\"{o}}ck, and
  Vidal]{Daley2004}
Daley,~A.~J.; Kollath,~C.; Schollw{\"{o}}ck,~U.; Vidal,~G. {Time-dependent
  density-matrix renormalization-group using adaptive effective Hilbert
  spaces}. \emph{Journal of Statistical Mechanics: Theory and Experiment}
  \textbf{2004}, \relax
\mciteBstWouldAddEndPunctfalse
\mciteSetBstMidEndSepPunct{\mcitedefaultmidpunct}
{}{\mcitedefaultseppunct}\relax
\EndOfBibitem
\bibitem[Valentine(2020)]{Valentine2020b}
Valentine,~D. \emph{Singlet fission in linear $\pi$-conjugated systems}; DPhil
  thesis, University of Oxford: Oxford, 2020\relax
\mciteBstWouldAddEndPuncttrue
\mciteSetBstMidEndSepPunct{\mcitedefaultmidpunct}
{\mcitedefaultendpunct}{\mcitedefaultseppunct}\relax
\EndOfBibitem
\bibitem[Bursill and Barford(1999)Bursill, and Barford]{Bursill1999}
Bursill,~R.~J.; Barford,~W. {Electron-lattice relaxation, and soliton
  structures and their interactions in polyenes}. \emph{Physical Review
  Letters} \textbf{1999}, \emph{82}, 1514--1517\relax
\mciteBstWouldAddEndPuncttrue
\mciteSetBstMidEndSepPunct{\mcitedefaultmidpunct}
{\mcitedefaultendpunct}{\mcitedefaultseppunct}\relax
\EndOfBibitem
\bibitem[Barford \latin{et~al.}(2002)Barford, Bursill, and
  Lavrentiev]{Barford2002c}
Barford,~W.; Bursill,~R.~J.; Lavrentiev,~M.~Y. {Breakdown of the adiabatic
  approximation in trans-polyacetylene}. \emph{Physical Review B - Condensed
  Matter and Materials Physics} \textbf{2002}, \emph{65}, 1--5\relax
\mciteBstWouldAddEndPuncttrue
\mciteSetBstMidEndSepPunct{\mcitedefaultmidpunct}
{\mcitedefaultendpunct}{\mcitedefaultseppunct}\relax
\EndOfBibitem
\bibitem[Bursill and Barford(2002)Bursill, and Barford]{Bursill2002}
Bursill,~R.~J.; Barford,~W. {Large-scale numerical investigation of excited
  states in poly(para-phenylene)}. \emph{Physical Review B - Condensed Matter
  and Materials Physics} \textbf{2002}, \emph{66}, 1--8\relax
\mciteBstWouldAddEndPuncttrue
\mciteSetBstMidEndSepPunct{\mcitedefaultmidpunct}
{\mcitedefaultendpunct}{\mcitedefaultseppunct}\relax
\EndOfBibitem
\bibitem[Bursill and Barford(2009)Bursill, and Barford]{Bursill2009}
Bursill,~R.~J.; Barford,~W. {Symmetry-adapted density matrix renormalization
  group calculations of the primary excited states of poly(para-phenylene
  vinylene)}. \emph{Journal of Chemical Physics} \textbf{2009},
  \emph{130}\relax
\mciteBstWouldAddEndPuncttrue
\mciteSetBstMidEndSepPunct{\mcitedefaultmidpunct}
{\mcitedefaultendpunct}{\mcitedefaultseppunct}\relax
\EndOfBibitem
\end{mcitethebibliography}

\end{document}